\documentclass[aps,prd,10pt,nofootinbib,twocolumn,superscriptaddress,floatfix,notitlepage]{revtex4-1}
\pdfoutput=1

\usepackage{graphicx}
\usepackage{amsmath,amsfonts,amssymb}
\usepackage[dvipsnames]{xcolor}
\usepackage[breaklinks,colorlinks,urlcolor=MidnightBlue,citecolor=WildStrawberry,linkcolor=Fuchsia]{hyperref}
\usepackage{verbatim}
\usepackage{enumitem}
\usepackage{aas_macros}
\usepackage{subcaption}

\newcommand{\td}{{\rm d}}

\newcommand{\be}{\begin{equation}}
\newcommand{\ee}{\end{equation}}
\newcommand{\bea}{\begin{equation} \begin{aligned}}
\newcommand{\eea}{\end{aligned} \end{equation}}

\def\lsim{\mathrel{\raise.3ex\hbox{$<$\kern-.75em\lower1ex\hbox{$\sim$}}}}
\def\gsim{\mathrel{\raise.3ex\hbox{$>$\kern-.75em\lower1ex\hbox{$\sim$}}}}

\usepackage[capitalise]{cleveref}
\usepackage[activate={true,nocompatibility},final,kerning=true,factor=1100,stretch=10,shrink=10]{microtype}
\usepackage{academicons}
\usepackage{fontawesome5}
\definecolor{orcidlogocol}{rgb}{0.65, 0.807, 0.223}
\newcommand{\orcid}[1]{$\,$\href{https://orcid.org/#1}{\textcolor{orcidlogocol}{\faOrcid}}}

\begin{document}

\title{The stochastic gravitational wave background from cosmic superstrings}

\author{Anastasios Avgoustidis\orcid{0000-0001-7247-5652}}\email{anastasios.avgoustidis@nottingham.ac.uk}

\author
{Edmund J. Copeland\orcid{0000-0003-3959-6051}}\email{edmund.copeland@nottingham.ac.uk}

\author{Adam Moss\orcid{0000-0002-7245-7670}}\email{adam.moss@nottingham.ac.uk}

\author{Juhan Raidal\orcid{0009-0002-2600-3632}}
\email{juhan.raidal@nottingham.ac.uk}
\affiliation{School of Physics and Astronomy, The University of Nottingham, Nottingham, NG7 2RD, UK}

\begin{abstract}
We study the stochastic gravitational wave background sourced by a network of cosmic superstrings and demonstrate that incorporating higher-mass string species, beyond the fundamental string, is crucial for accurately modeling the resulting gravitational wave spectrum across frequencies ranging from nanohertz to kilohertz.
Using the multi-tension velocity-dependent one-scale model to evolve the cosmic superstring network, we perform several fits to the NANOGrav 15-year dataset and obtain expectation values for the fundamental string tension, string coupling and effective size of compact extra dimensions. We find that the cosmic superstring best-fits are comparable in likelihood to Supermassive Black Hole models, thought by many to be the leading candidate explanation of the signal. The implications of the best-fit spectra are discussed within the context of future gravitational wave experiments. We obtain expectation values for the fundamental string tension of $\log_{10}(G\mu_1)=-11.4^{+0.3}_{-0.2}$($-11.5^{+0.3}_{-0.2}$) for gravitational waves originating from large cuspy (kinky) cosmic superstring loops and $\log_{10}(G\mu_1)=-9.7^{+0.7}_{-0.7}$($-9.9^{+1.0}_{-0.5}$) for small cuspy (kinky) loops. We also place $2\sigma$ upper bounds on the string coupling, finding $g_s<0.7$ in all cases, and comment on the implication of our results for the effective size of the compact extra dimensions.
\end{abstract}

\maketitle

\section{Introduction}

Pulsar timing arrays (PTAs) have provided mounting evidence towards the existence of a stochastic gravitational wave background (SGWB) in the nanohertz frequency range. This is indicated by both the measurement of a common-spectrum stochastic process and recent strong evidence of Hellings-Downs angular correlations in the measured signal \cite{Hellings:1983fr,NANOGrav:2023gor,Reardon:2023gzh,Zic:2023gta,EPTA:2023fyk,Xu:2023wog}. Since the announcement of the evidence, there have been a number of explanations put forward for the source of the SGWB. Probably the favourite to date arises from astrophysics; a population of inspiralling supermassive black hole binaries with some additional low-frequency energy loss mechanisms, such as large eccentricity or interactions with the environment \cite{NANOGrav:2023gor,Ellis:2023dgf,EPTA:2023xxk,Bi:2023tib,Ellis:2024wdh,Raidal:2024odr}. With the current state of measurements, however, many new physics inspired cosmological sources, such as inflation, domain walls, cosmic superstrings, scalar induced gravitational waves, early-universe phase transitions and others \cite{NANOGrav:2023hvm,Figueroa:2023zhu} have been seen to fit the data equally well, if not better. In fact, the only cosmological model considered by the NANOGrav collaboration when analyzing their 15-year dataset (NG15) which did not provide a comparable fit to the supermassive black hole model was cosmic strings, although one of the best fits to the NG15 data was a model of cosmic superstrings \cite{NANOGrav:2023hvm,Figueroa:2023zhu,Ellis:2023tsl}.

Cosmic superstrings occur in certain string theory scenarios where fundamental strings (F-strings) and one-dimensional Dirichlet branes (D-strings) are stretched to cosmic scales by brane inflation \cite{Sarangi:2002yt,Jones:2002cv,Majumdar:2002hy,Copeland:2003bj}. Unlike ordinary cosmic string networks, which generally consist of a single type of string with unit chopping probability when intersecting, cosmic superstrings have multiple different species (the F- and D-strings together with their bound states) and a reduced probability of intersecting. As they are string theory objects, the extra dimensions they experience along with their more complicated dynamics\footnote{Note that non-trivial features like multiple species and reduced intercommutation probabilities can also arise in different contexts, notably in pure Yang-Mills theory \cite{Yamada:2022imq,Yamada:2022aax}.} significantly increases the difficulty of modelling the evolution of these networks. Nevertheless, remarkable strides have been made towards a consistent picture of cosmic superstring network evolution by modifying methods used to study cosmic strings \cite{Avgoustidis:2009ke,Avgoustidis:2007aa,Pourtsidou:2010gu,Marfatia:2023fvh,Correia:2022spe,Blanco-Pillado:2021ygr}. While there remain uncertainties in the detailed behaviour and interactions of cosmic superstring  networks, we believe that we are at a point where detailed predictions can be made about the spectrum of stochastic gravitational waves arising from them \cite{Sousa:2016ggw}.

In this paper, we present an overview of the most up-to-date modelling techniques for a network of multi-tension cosmic superstrings, along with a calculation of the SGWB generated by them, fitting this background to the NANOGrav 15-year dataset. The outline of the paper is as follows. In section \ref{sec:model}, we discuss the model used to describe the dynamics of the network and the gravitational waves (GWs) generated by it. In section \ref{sec:behaviour} we show how the dynamics and SGWB depend on model parameters and specify the parameter choices made for fitting to the NG15 data. In section \ref{sec:results}, we show fits to the NG15 data for different sizes of cosmic superstring loops and for different loop structures and discuss their implications before concluding in section \ref{sec:conclusions}.

\section{Modelling}
\label{sec:model}

\subsection{String Dynamics}

Cosmic superstring networks consist of light fundamental F-strings with tension $\mu_{\rm F}$, heavy D-strings with non-perturbative tension $\mu_{\rm D}=\mu_{\rm F}/g_s$, where $g_s<1$ is the fundamental string coupling, and heavier bound states between $p$ F-strings and $q$ D-strings, with $p$ and $q$ coprime. They are collectively referred to as $(p,q)$-strings, with each species carrying a conserved charge $(p,q)$ and having string tension
\bea \label{pg-string-tension}
    \mu_{(p,q)}=\frac{\mu_{\rm F}}{g_s}\sqrt{p^2 g_s^2+q^2}\, .
\eea

We model a population of cosmic superstrings as Nambu-Goto strings using the multi-tension string network velocity-dependent one-scale (VOS) model \cite{Avgoustidis:2009ke,Avgoustidis:2007aa,Pourtsidou:2010gu}. To simplify the notation, we label each pair of charges $(p_i,q_i)$ by a single index $i$, so that the $i^{\rm th}$ string species has tension $\mu_i\equiv \mu_{(p_i,q_i)}$. We choose the ordering so that the lightest F-string species, carrying charge $(1,0)$, have tension $\mu_1\equiv \mu_{\rm F}$, the D-strings with charge (0,1) have tension $\mu_2\equiv \mu_{\rm D}$, the $1^{\rm st}$ bound state (the FD-string) with charge $(1,1)$ has tension $\mu_3\equiv \frac{\mu_{\rm F}}{g_s}\sqrt{1+g_s^2}$, and so on.      

As the name suggests, the VOS approach describes each species of string, $i$, in the network through a single characteristic length scale, the correlation length, $L_i$. String dynamics are introduced into the model via the root-mean-square velocity, $v_i$, of each string. As the string network evolves, the different types of strings interact and intercommute with different probabilities. The cosmological evolution of such a multi-tension string network is described by the set of equations \cite{Avgoustidis:2007aa,Avgoustidis:2009ke,Pourtsidou:2010gu}
\bea\label{eq:VOSL}
    \frac{\td L_i}{\td t}&=H(t) L_i (1+v_i^2)+\frac{1}{2}\Tilde{c}_i v_i\\
    &+\frac{1}{2}\bigg(\sum_{a,k}\Tilde{d}^k_{ia}\frac{v_{ia}\ell_{ia}}{L_i^2L_a^2}-\sum_{b,a\leq b}\Tilde{d}^i_{ab}\frac{v_{ab}\ell_{ab}}{L_a^2L_b^2}\bigg)L_i^3\, ,
\eea
\bea\label{eq:VOSv}
    \frac{\td v_i}{\td t}=(1-v_i^2)\bigg[&\frac{k(v_i)}{L_i} -2H(t) v_i \\
     + &B\sum_{b,a\leq b} \Tilde{d}^i_{ab}\frac{v_{ab}}{v_i} \frac{\mu_a+\mu_b+\mu_i}{\mu_i}\frac{\ell_{ab} L_i^2}{L_a^2L_b^2}\bigg]\, ,
\eea
where $H(t)$ is the Hubble parameter, $\tilde{c}_i$ is the {\it self-intersection} (or loop chopping) efficiency parameter for strings of type $i$, and $\Tilde{d}^k_{ij}=\Tilde{d}^k_{ji}$ are the {\it cross-string intersection} efficiency parameters describing processes in which a string of type $i$ interacts with a string of type $j$, with relative velocity $v_{ij}$, to produce a type $k$ string segment of length $\ell_{ij}$. Such cross-string interactions can be thought of as ``zipping" processes in which the colliding strings join together along part of their length to produce a ``zipper", thus forming a Y-type junction, i.e. a trilinear vertex on which the three strings meet. As the tension of the zipper string is smaller than the sum of the tensions of the two colliding strings (which is required for the interaction to be possible; otherwise the corresponding coefficient $\tilde{d}^k_{ij}$ is zero), the formation of the junction/zipper liberates energy. The parameter $0\leq B\leq 1$ is introduced in the model to describe the fraction of this energy that is redistributed back into the string network. In particular, for $B=0$, all the energy released is assumed to be radiated away, while for $B=1$ all that energy is redistributed back into the network as kinetic energy of the formed zipper. 

The momentum parameter\footnote{In the original formulation of the VOS model \cite{Martins:1996jp} the momentum parameter was a constant, so that the model had two free parameters, namely the momentum parameter $k$ and the loop chopping efficiency $\tilde c$, which took different values in the matter and radiation eras. However, in \cite{Martins:2000cs} the momentum parameter was promoted to a function of $v$ so that the model had only one free parameter, $\tilde c$, and it was found that a single value $\tilde c=0.23 \pm 0.04$ provides an excellent fit to numerical simulations throughout cosmological evolution.}
\bea\label{eq:kofv}
    k(v_i)=\frac{2\sqrt{2}}{\pi}(1-v_i^2)(1+2\sqrt{2}v_i^3)\frac{1-8v_i^6}{1+8v_i^6}
\eea
describes the effective curvature of the $i^{\rm th}$ string network component and indirectly encodes the effect of small-scale structure on the acceleration of string segments at the scale of the correlation length \cite{Martins:2000cs}. 
As the root-mean-square velocities of cosmic (super)strings generally have mildly relativistic values, we approximate the average relative velocity between colliding string segments of type $i$ and $j$ by the (non-relativistic) result $v_{ij}=\sqrt{v_i^2+v_j^2}$. For the average length of the zipper segment resulting from the collision of two string segments of types $i$ and $j$ we take $\ell_{ij}^{-1}=L_i^{-1}+L_j^{-1}$, a value smaller than, but close to, the smallest of the two colliding segment lengths.

For cosmic superstrings, the self- and cross- interaction coefficients, $\Tilde{c}_i$ and $\Tilde{d}^k_{ij}$, in the multi-tension VOS model can be related to a microphysical intercommuting probability ${\mathcal P}_{ij}$, which can be computed (or, for interactions involving only heavy non-perturbative strings, approximated) using string theory techniques \cite{Jackson:2004zg,Hanany:2005bc}. As strings are not perfectly straight and have non-trivial small-scale-structure, two colliding segments have more than one opportunity to intercommute in each crossing time, and so the effective interaction coefficients have a weaker than linear dependence on the microphysical intercommutation probability. Following \cite{Pourtsidou:2010gu}, we take 
\bea\label{eq:effinterc}
    \Tilde{c_i}=\Tilde{c}\times\mathcal{P}_i^{1/3}\ \ ,\ \ \ \Tilde{d}_{ij}= \mathcal{P}_{ij}^{1/3}\, ,
\eea
where $\mathcal{P}_i \equiv \mathcal{P}_{ii}$ is the microphysical self-interaction probability for strings of type $i$. The prefactor $\Tilde{c}$ ensures that for ${\mathcal P}_{i}=1$ one recovers the standard VOS value $\Tilde c$ \cite{Martins:2000cs} for the loop chopping efficiency, while the power $1/3$ is supported \cite{Avgoustidis:2005nv} by numerical simulations of Nambu-Goto strings with a suppressed intercommutation probability, in both matter and radiation dominated cosmologies\footnote{ Earlier simulations of Nambu-Goto strings evolving in Minkowski spacetime may suggest a slightly stronger dependence  \cite{Sakellariadou:1990nd,Dvali:2003zj} of $\Tilde c$ on the intercommuting probability $P$, namely $\tilde c\propto P^{1/2}$.}. Note that going from $\Tilde{d}_{ij}$ in equation (\ref{eq:effinterc}) to $\Tilde{d}^k_{ij}=\Tilde{d}_{ij}S^k_{ij}$ in equations (\ref{eq:VOSL}-\ref{eq:VOSv}) involves a kinematic factor $S^k_{ij}$ \cite{Avgoustidis:2009ke,Pourtsidou:2010gu} (also see \cite{Copeland:2006if,Salmi:2007ah,Copeland:2007nv}) quantifying the relative probability that the collision proceeds via the additive or subtractive channel \cite{Tye:2005fn}, i.e. $(p,q)+(p',q')\longrightarrow (p\pm p', q\pm q')$, producing a heavier or lighter string segment, respectively.        

The microphysical intercommutation probability ${\mathcal P}_{ij}$ deserves closer attention. It encodes both the quantum nature of cosmic superstring interactions \cite{Jackson:2004zg,Hanany:2005bc} and the effect of compact extra dimensions, which introduce a suppression in the intercommutation probability as the colliding strings can miss each other in the extra dimensions \cite{Jones:2003da,Dvali:2003zj,Jackson:2004zg}. For fixed $i$ and $j$ the value of ${\mathcal P}_{ij}$ depends on the relative velocity $v_{ij}$ between the colliding strings, the collision angle $\theta$, the fundamental string coupling $g_s$ and the compactification characteristics of the theory. However, in constructing the effective intercommutation coefficients for the VOS model, one integrates over collision angles and relative velocities, and so the resulting $\Tilde{c_i}$ and $\Tilde{d}^k_{ij}$ only depend on $g_s$ and the compactification. This introduces the exciting possibility of constraining fundamental parameters of string theory using CMB \cite{Avgoustidis:2011ax, Charnock:2016nzm} or GW \cite{Sousa:2016ggw} data.  

While the dependence on $g_s$ is relatively straightforward, the impact of extra dimensions on ${\mathcal P}_{ij}$ (and, in turn, on the VOS parameters $\Tilde{c_i}$, $\Tilde{d}^k_{ij}$) is more subtle. Following \cite{Pourtsidou:2010gu} we parametrise the effects of extra dimensions through a single volume parameter $w$, defined \cite{Jackson:2004zg} as the ratio of the minimum volume, $V_{\rm min}$, that can be occupied by the string in the extra dimensions (this can be roughly thought of as the string thickness raised to the number of compact extra dimensions \cite{Dvali:2003zj}), over the total volume available to the fundamental strings, $V_{\rm FF}$, which is bounded above by the volume of the compact dimensions. Thus, $w$ can be thought of as the effective size of the compact extra dimensions. A small value of $w$ means that the strings can explore a large volume in comparison to the string thickness, while $w=1$ corresponds to the extra dimensions being compactified at the string scale, or being strongly warped such that the strings are stabilised in a small region (of size comparable to the string thickness) of the compactification manifold. Therefore, for $w=1$ extra-dimensional effects can be ignored and the intercommutation probabilities are determined solely by the quantum mechanical interactions between the different types of strings (i.e. there is no additional volume suppression due to the extra dimensions). Note that because heavy non-perturbative strings fluctuate less than lighter strings, they see a smaller effective volume and so compactification effects are less important for heavy strings than for the light F-strings. For more details on $w$ and its interpretation, see \cite{Pourtsidou:2010gu,Jackson:2004zg,Jackson:2007hn}.

To summarise, given a choice for the parameter $B$ and the relations between $\ell_{ij}$ and $v_{ij}$ to the model's dynamical variables, $v_i$ and $L_i$ respectively, the multi-tension VOS model for cosmic superstrings depends on three free parameters: the fundamental string tension $\mu_1\equiv \mu_{\rm F}$, the string coupling $g_s$, and the effective volume parameter $w$.     

In cosmic superstring networks, there are, in principle, an infinite number of string species carrying all possible coprime pairs of charges $(p,q)$. In practice, however, the first few lightest string species are the most numerous and dominate the energy density of the network \cite{Pourtsidou:2010gu}. This has, in some cases, led to an oversimplification where it has been assumed that modelling the evolution of just the lightest string type is sufficient to describe the dynamics of the system. In such an approximation, one models the network through a single correlation length (and rms velocity) but also introduces a suppressed probability for string intercommutation. The model then has two free parameters, namely the string tension and the intercommuting probability. As we will show in \ref{sec:stringcomparison}, while this might be a good approximation for some areas of the parameter space, it is not an accurate description in general.

\subsection{Loop Number Density} \label{subsec:LND}

During the evolution of the cosmic superstring network, strings interact and intersect, forming string loops of different charges and tensions. These loops oscillate relativistically and decay by emitting GWs, giving rise to a SGWB. It is then useful to consider the string network of each string species, $i$, as the sum of two components; a {\it long string network} characterised by the correlation length $L_i$ and root-mean-square velocity $v_i$ satisfying equations (\ref{eq:VOSL}-\ref{eq:VOSv}) and a {\it loop network} consisting of loops with a certain length distribution.     

To study the GW signature of cosmic superstrings, it is crucial to determine the number densities of the associated loops. For each string type $i$, the energy density lost into loops from the long-string network is given by
\bea
    \Dot{\rho}_{\text{loop},i}=\frac{\Tilde{c}_i v_i}{\gamma(v_i)L_i}\rho_i\, ,
\eea
where $\rho_i \equiv \mu_i/L_i^2$ is the energy density of the long string network for each string species, $\gamma(v_i)=(1-v_i^2)^{-1/2}$ accounts for the redshifting of loop velocities \cite{Vilenkin:2000jqa} of species $i$, and $\Tilde{c}_i$ are the corresponding self-interaction coefficients.

Following \cite{Pourtsidou:2010gu}, the string self-interaction coefficients, $\Tilde{c}_i$, are calculated via
equation (\ref{eq:effinterc}), where $\tilde{c}$ is extracted from numerical Nambu-Goto (single-tension) network simulations and has a value $\tilde{c}=0.23\pm 0.04$ throughout\footnote{It is also common to use different values for the loop chopping efficiency parameter in different eras, namely $\tilde{c}_r=0.23$ in the radiation era and $\tilde{c}_m=0.18$ in the matter era \cite{Vilenkin:2000jqa,Martins:1996jp}, and introduce an interpolating function \cite{Battye:1997hu} for the radiation to matter transition. However, this is not necessary when one uses the velocity dependent momentum parameter (\ref{eq:kofv}) and both approaches produce practically identical results.} cosmological evolution \cite{Martins:2000cs}.

Note that, during the transition to the matter era and beyond, it is important to include the full cosmological time-dependence in both the scale factor $a(t)$ and in $\mathcal{P}_i$ (calculated from the string velocities which depend on the Hubble parameter). 

Although the string velocities quickly reach constant values in the radiation era and the scale factor is well-approximated by assuming radiation domination in the Friedmann equation, this is not true for the matter era, which transitions to the cosmological constant-dominated regime soon after the network enters matter scaling.

Since $\rho_{\text{loop},i}$ must match the energy density in the loops of species $i$ at all times $t$,
\bea
    \rho_{\text{loop},i}=\int \mu_i l \times n_i(l,t) \td l\, ,
\eea
it can be shown that the number density of type-$i$ string loops with lengths between $l$ and $l+\td l$ at time $t$ is \cite{NANOGrav:2023hvm}
\bea\label{eq:nloop}
    n_i(l,t)&=\frac{\Tilde{c}_i v_i(t_b)}{\gamma(v_i(t_b))\xi_i^3(t_b)}\\
    &\times\frac{\Theta(t-t_b)\Theta(t_b-t_{i})}{\alpha_i(t_b)(\alpha_i(t_b)+\Gamma G\mu_i + \Dot{\alpha_i}(t_b)t_b)t_b^4}\frac{a(t_b)^3}{a(t)^3}\, ,
\eea
where, at birth $t_b$, loops have a length of $l_i(t_b)=\alpha_i(t_b)t_b,$ where $\alpha_i(t_b)$ is usually taken to be a constant, and $\xi_i=L_i/t$ is the time-normalized correlation length. The coefficient $\Gamma$ reflects the gravitational power emitted in the string network, $\Gamma G\mu_i$, for species $i$ and is typically taken to be $ \Gamma \sim 50$ \cite{Blanco-Pillado:2017oxo,quashnock_gravitational_1990} (this is discussed further in Sec. \ref{sec:gamma}). The Heaviside theta functions ensure that only loops born before the current time and after some initial time $t_{i}$ are counted.

It is important to discuss some conventions regarding loop sizes which can cause confusion. Loop sizes are commonly described as a fraction of some physical scale in the system and this fraction is usually denoted by $\alpha$. Common examples include $l=\alpha_h d_h=\alpha_L L=\alpha t$ (here we have suppressed the species index), where $d_h$ is the horizon size at a given time. It is also, unfortunately, common to drop the subscript of different $\alpha$, making it easy to mistake one for the other. For the rest of this paper, we will be using $\alpha_i=l_i/t$ for each species $i$, unless  explicitly stated otherwise.

The exact values are unknown for cosmic superstrings, but numerical simulations of cosmic gauge strings can provide some hints towards the magnitude of $\alpha_i$. This is discussed in detail in section~\ref{sec:loopsize}.

\subsection{Gravitational Wave Emission}
\label{sec:GWemission}

Due to the relativistic oscillation speeds of string loops, they provide the dominant energy contribution to GWs arising from a cosmic superstring network. Emission from a large number of loops chopped from the string network throughout cosmological evolution results in a superposition of GWs -- a SGWB. A loop of length $l_i$ emits GWs with frequencies
\bea \label{emitted-freq}
    f_{j,i}=\frac{2j}{l_i}\,
\eea
in the $j^{\rm th}$ harmonic mode. This emission decreases the loop size such that the length at time $t$ (in terms of the time of its birth $t_b$) is
\bea
    l_i(t)= l_i(t_b)-\Gamma G\mu_i(t-t_b)\, .
\eea
It follows that the time of birth of a string loop of length $l_i$  can be found by solving \cite{Vilenkin:2000jqa,NANOGrav:2023hvm}
\bea\label{eq:tb}
    t_b(l_i,t)=\frac{l_i + \Gamma G\mu_i t}{\alpha_i(t_b) + \Gamma G\mu_i}\,.
\eea

The GW power emitted in the $j^{\rm th}$ harmonic mode is given by \cite{Caprini:2018mtu}
\bea \label{harmonic-power}
    \frac{\td E_j}{\td t}=G\mu^2\frac{\Gamma}{\mathcal{E}}j^{-q}\, ,
\eea
where $q$ depends on the structure of the emitting loops, and $\mathcal{E}\equiv \sum_j^{j_{\rm max}}j^{-q}$ represents the fact that because the higher frequency modes (higher harmonics) are damped by gravitational back-reaction, the full GW power obtained by a summation over harmonics does not need to include them all to accurately model the GW signal; rather it can be cut off at a maximum harmonic mode $j_{\rm max}$.

For emission dominated by loop cusps, points on the loop which instantaneously move at the speed of light giving rise to a beamed burst of gravitational radiation, $q=4/3$. For emission dominated by kinks, discontinuities in the loop tangent vector moving along the loop at the speed of light, $q=5/3$. Due to the stronger signal in Eq.~(\ref{harmonic-power}), GW emission is usually assumed to be dominated by cusps. The validity of this assumption is discussed in section~\ref{sec:cuspsVskinks}.

The energy density of the GW background radiation sourced by loops of cosmic superstring is given, per logarithmic frequency interval (and in units of critical density), by \cite{Caprini:2018mtu}
\bea
    \Omega_{\rm GW}(f)=\frac{1}{\rho_c}\frac{\td \rho_{\rm GW}}{\td \ln{f}}\, ,
\eea
where the critical density today is $\rho_c=3H_0^2/(8\pi G)$ and
\bea
    \rho_{\rm GW}=\sum_i^{N}\rho_{\rm{GW},\textit{i}}\, ,
\eea
with $\rho_{\rm{GW},\textit{i}}$ the gravitational wave energy density emitted by each of the $N$ string species.

In terms of GW energy emitted by each harmonic mode, the total energy density can be written as \cite{Sousa:2016ggw} 
\bea\label{eq:Omegajsum}
    \Omega_{\rm GW}(f)=\sum_i^N\sum_j^{j_{\rm max}}\frac{j^{-q}}{\mathcal{E}}\Omega_{\rm GW,\textit{i}}^j(f)
\eea
and
\bea\label{eq:omegaj}
    \Omega_{\rm GW,\textit{i}}^j(f)=\frac{16\pi}{3}\bigg(\frac{G\mu}{H_0}\bigg)^2\frac{j\Gamma}{f}\int_{t_i}^{t_0} n_i(l_j(t),t)a(t)^5 \td t\, .
\eea
Here, $t_i$ is the time at which loop production becomes significant and $n_i(l,t)$ is the loop number density given by Eq.~(\ref{eq:nloop}). We assume that the initial time is the end of the friction-dominated regime for single-string networks, $t_i\approx t_{\rm Pl}/(G\mu)^2$, where $t_{\rm Pl}$ is the Planck time \cite{Vilenkin:2000jqa}.

The integral in Eq. (\ref{eq:omegaj}) sums over all loops with lengths
\bea \label{length-at-t}
    l_j(t)=\frac{2j}{f}a(t)\,,
\eea
which is the physical length of loops at time $t$ that emit GWs (in their $j^{\rm th}$ harmonic) with a measured frequency of $f$ at the present time.

The summation over harmonic modes in Eq.~(\ref{eq:Omegajsum}) is made much faster computationally by noting that in Eq.~(\ref{eq:omegaj}), we can make the formal identification between the $i^{\rm th}$ string species emitting with harmonic $j$ at frequency $f$, and that emitting with harmonic $1$ but at frequency $f/j$, namely $\Omega_{\rm GW,\textit{i}}^j(f)=\Omega_{\rm GW,\textit{i}}^1(f/j)$ which then implies that above some large harmonic number, $M$, the spectrum of harmonics becomes nearly continuous such that \cite{Ellis:2023tsl,Cui:2019kkd}
\bea
    \Omega_{\rm GW\textit{i}}^j(f)\approx \sum_{j=1}^M \Omega_{\rm GW,\textit{i}}^1(f/j)+\int_{M+1}^{j_{\rm max}} \Omega_{\rm GW,\textit{i}}^1(f/j) \td j\, .
\eea
Following \citet{Ellis:2023tsl}, we set $M=1000$.

\section{Stochastic gravitational wave background spectra}
\label{sec:behaviour}

\subsection{Importance of heavier strings}
\label{sec:stringcomparison}

As mentioned in section~\ref{subsec:LND}, the energy density in a network of infinitely long strings of species $i$, where $L_i$ is the correlation length of the long string, is given by,
\bea\label{eq:rho}
    \rho_i=\frac{\mu_i}{L_i^2}.
\eea
We are considering a population of many string species, with fundamental strings having a much smaller correlation length than those of the heavier string species. Hence their energy density dominates the system. This fact has been used to make the approximation that a network of cosmic superstrings can be modelled as a single-string network with a reduced interaction probability (essentially rescaling the GW spectrum produced by a network of Nambu-Goto cosmic strings). Although this approximation may be valid for some specific parameter values and frequency ranges, it was shown by \citet{Sousa:2016ggw} that there are characteristic features in the GW emission from cosmic superstring networks that cannot be captured by a single string species model. In particular, if the loops chopped from the network are small ($\alpha_i\sim 2\Gamma G \mu_i$), a distinct multi-peaked signature in the spectrum is seen at low frequencies, just below the nano-Hz region\footnote{Note, however, that \cite{Sousa:2016ggw} work in terms of $\alpha_L$ so the relation between the physical loop size and $\mu_i$ in that paper also involves a factor of the time normalised correlation length $\xi_i$.}. 

Although \cite{Sousa:2016ggw} only investigated a model with three species of superstring, as more and more string species are considered, we should see a cascade of low-frequency peaks emitted by heavier and heavier string species. However, as the correlation length of the strings increases as the string species become heavier, the amplitude of the associated GWs emitted decreases. It is not immediately clear how many string species should be included before additional GW contributions become negligible. Furthermore, even if heavier strings themselves do not emit a measurable amount of GWs, they can still impact the evolution of the lighter strings and need to be included when solving the VOS equations.

To look at the impact of heavier string species, we pick a benchmark model where all intersection energy between strings is radiated away ($B=0$) and the extra dimensions are compactified close to the string thickness scale, or are highly warped ($w=1$). The effect of varying these parameter values is investigated in section~\ref{sec:otherParams}. Following \cite{Pourtsidou:2010gu}, we truncate the number of string species considered at $N=7$ lightest strings, with corresponding charges
\bea
    {(p_i,q_i)}={(1,0),(0,1),(1,1),(2,1),(1,2),(3,1),(1,3)}\, ,
\eea
where $i=1,..,7$ and the tensions $\mu_i=\mu_{(p_i,q_i)}$ are given by Eq.~(\ref{pg-string-tension}).

In the context of GW backgrounds, previous work has focused on only including the three lightest (F, D and FD) strings \cite{Sousa:2016ggw,Marfatia:2023fvh}. However, even if the GW signal from heavier strings is negligible, more than three species must be included in the network evolution in order to accurately predict the correlation lengths (or, equivalently, the number densities) of the three lightest species. For example, collisions between the lightest (and most abundant) $i=1$ strings and the heavy (and relatively scarce) $i=4$ strings predominantly lead to the production of $i=3$ strings via the dominant subtractive channel $(1,0)+(2,1)\longrightarrow (1, 1)$, which can significantly affect the number density of the $i=3$ FD strings. To illustrate the impact of including heavier strings on the evolution of the three lightest species, we solve the VOS equations Eqs.~(\ref{eq:VOSL})-(\ref{eq:VOSv}) for both $i=1,\ldots,3$ (3-string model) and $i=1,\ldots,7$ (7-string model).

Instead of looking at the correlation lengths of each species, it is more informative to look at the normalized characteristic length $\xi_i(t)=L_i(t)/t$. This factors out the increase in correlation length due to the expansion of the universe and gives a clearer picture of the dynamics of the string network. For example, the network spends much of its time in a scaling regime, where the correlation length is proportional to the horizon size, so that $L_i(t)\propto t$ and $\xi_i(t)$ is a constant.

\begin{figure*}
    \centering    
    \includegraphics[width=0.95\textwidth]{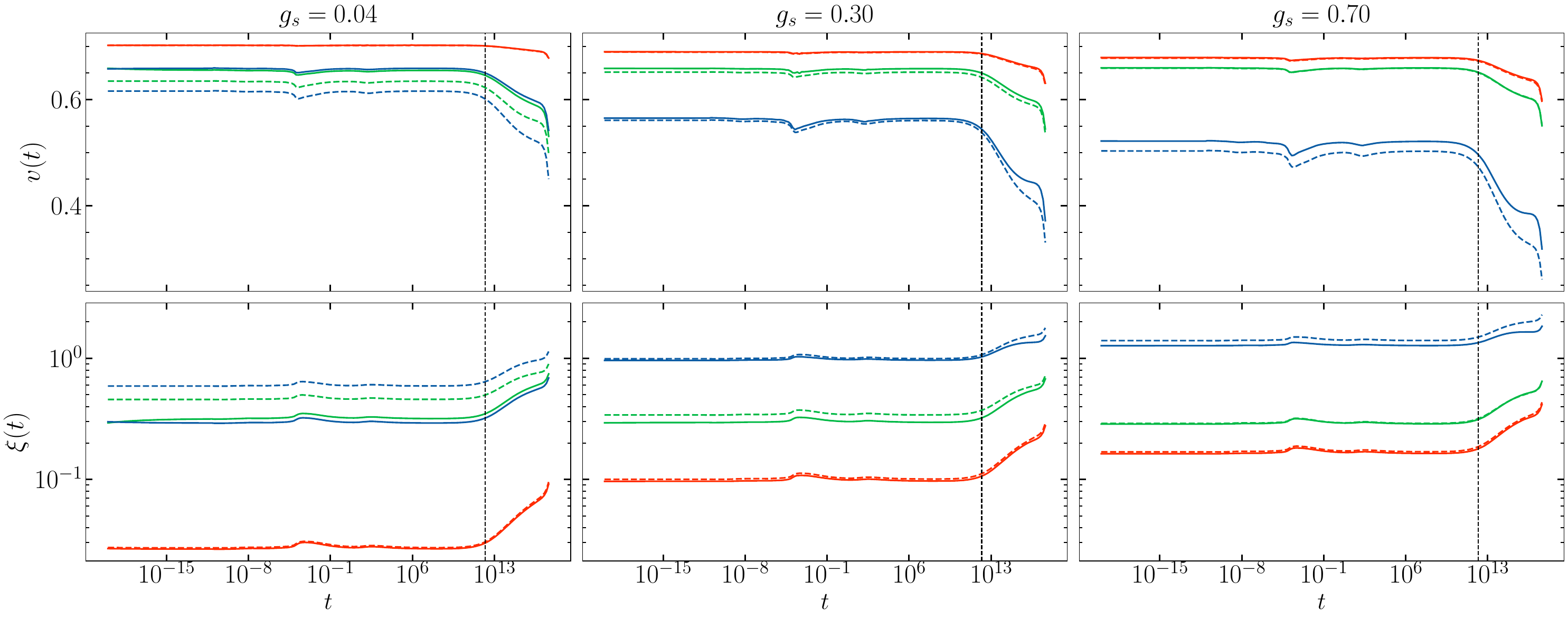}

    \caption{The correlation lengths and velocities of F-strings (red), D-strings (green) and FD-junctions (blue) for both the 7-string (solid lines) and 3-string (dashed lines) models at different values of the string coupling constant. The vertical dashed lines indicate the epoch of matter-radiation equality.}
    \label{fig:3vs7}
\end{figure*}

The root-mean-square speed and normalized characteristic lengths obtained by solving the VOS equations can be seen in Fig. \ref{fig:3vs7} for different values of the string coupling, $g_s$. Although the specific scaling values differ, the general characteristics of the speeds and lengths are the same in both models. During the radiation era, both parameters are very well approximated by a constant linear scaling regime value, only deviating from scaling during short periods following a change in relativistic degrees of freedom. During the transition to the matter era and at later times, no scaling behaviour is seen and both the velocities and correlation lengths vary continuously with time. This shows the importance of full modelling of the cosmological evolution, as it is commonly assumed that the string network spends most of its time in scaling 
during \textbf{both} the radiation and matter eras, which is not the case. Changes in the radiation-matter transition period have the strongest effect on the low-frequency end of the GW spectrum, which is also the most relevant when looking for GWs from heavier string species, making accurate modelling particularly important for cosmic superstring GW searches.

For all three string coupling values, there are clear differences between the 3-string and 7-string solutions. The general trend is that, as expected, the 3-string model predicts larger correlation lengths (smaller number densities) because it does not take into account heavier string species, which can decay to the three lightest ones. As strings of species $i\geq 4$ interact with and decay into lighter strings, the lighter string networks receive a boost of energy. This is seen as an increase in velocities and a corresponding decrease in correlation lengths. The size of these effects is dependent on the string coupling, with only the FD-string being affected at large couplings, whilst modifications to both the FD- and D-strings can be seen at low couplings. The fundamental string remains largely unaffected by the inclusion of heavier string species in the VOS equations. 

Although there are clear differences in the VOS parameters between the two models, since the dominant F-string behaves the same in both cases, it is unclear how much these differences will change the overall gravitational wave background spectrum. We now move on to discuss the effect of heavier string species on the GW signal from cosmic superstring networks.  In doing so, we will also justify that one only needs to consider the GWs coming from the three lightest string species, even through, as we have just seen, the seven lightest species must be included in the VOS modelling to accurately determine the network evolution of the three lightest strings. 

\subsubsection*{Impact of heavy species on the GW signal}
\label{sec:3v7}

As discussed in section~\ref{sec:stringcomparison}, the GW energy density emitted by a cosmic superstring network is dominated by its lightest species, except possibly at low frequencies where signatures from more than one species can appear. This does not change the fact that, due to the lower energy density arising from their larger correlation length in Eq. (\ref{eq:rho}), the magnitude of GWs seen from each heavier string species is significantly reduced. Indeed, one reason that heavier species can dominate the GW signal at low frequencies is that there is a sharp drop off of the signal as the frequency decreases below the peak, and so a heavier string, having a peak position at a lower frequency, can dominate the light string signal at low frequencies. As we will see shortly, this is not the complete picture, but it is clear that heavy string domination can only happen at amplitudes much lower than the light string GW peak amplitude, and each heavier species leads to a further suppressed GW signal (which can possibly dominate at low frequencies). 

Therefore, it is generally sufficient to only calculate GW contributions for the three lightest strings and, in the interest of computational time, we will be doing so throughout this paper. We stress that this does not mean that only three string species can be used when solving the VOS equations, as incorporating heavier ($i>3$) string species in modelling the network evolution leads to solutions of higher velocity and lower characteristic length than those obtained when only considering the three lightest strings (thus neglecting decays of heavier species), as seen in Fig. \ref{fig:3vs7}. We will now examine how this affects the GW background spectra generated by the cosmic superstring populations.

\begin{figure*}
    \centering
    \includegraphics[width=0.95\textwidth]{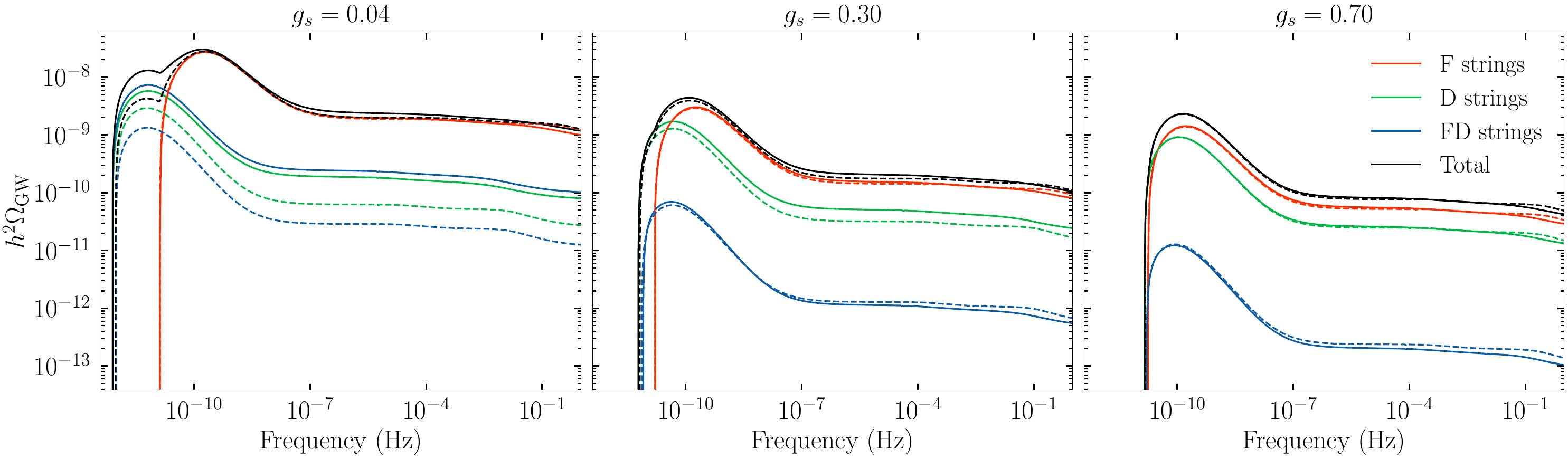}

    \caption{Plots showing the gravitational wave background spectra generated at different string coupling constants at string tension $G\mu_1=10^{-9}$ and loop size $\alpha_i=2\Gamma G\mu_i$. The solid lines show the GWs from the 7-string model and the dashed lines from the 3-string model.}
    \label{fig:smallLoops}
\end{figure*}

\begin{figure*}
    \centering  
    \includegraphics[width=0.95\textwidth]{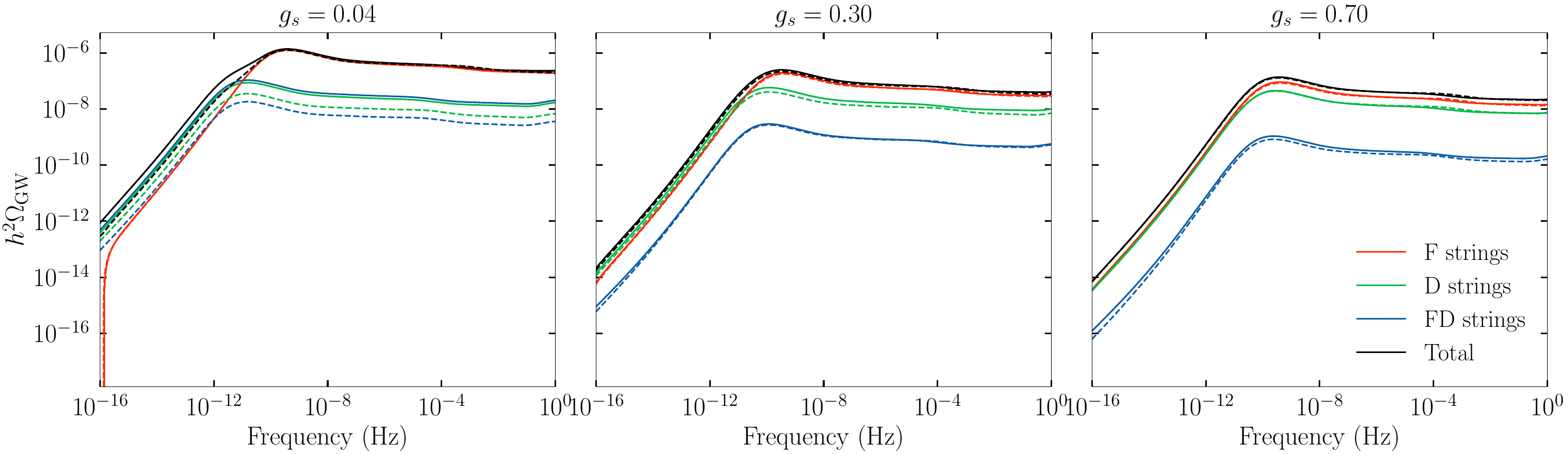}

    \caption{Plots showing the gravitational wave background spectra generated at different string coupling constants at string tension $G\mu_1=10^{-9}$ and loop size $\alpha_{L,i}=0.37$. The solid lines show the GWs from the 7-string model and the dashed lines from the 3-string model.}
    \label{fig:bigLoops} 
\end{figure*}

We generate spectra at two different scales for the loop size -- $\alpha_{L,i}\equiv l_i/L_i=0.37$ (large loops) and $\alpha_i\equiv l_i/t=2\Gamma G\mu_i$ (small loops). The exact choice of values will be discussed in section~\ref{sec:loopsize}, but these two scales will enable us to look at both the large and small loop behaviour of the spectra respectively. Initially, the other parameter values, ${G\mu_1=10^{-9},B=0,w=1,j_{\rm max}=1}$, are chosen to be identical to those in \cite{Sousa:2016ggw}, allowing us to make a direct comparison with their results. We will later allow for these parameters to vary. We find that changing them does not affect the nature of the differences between the 3-string and 7-string models.  We also choose to look at spectra generated at three different string coupling values, $g_s=0.04,\;0.3\;{\rm and}~0.7$, as the differences seen in Fig. \ref{fig:3vs7} were dependent on the coupling. The corresponding spectra calculated for the 3-string and 7-string models can be seen in Fig. \ref{fig:smallLoops} in the small loop regime and Fig. \ref{fig:bigLoops} in the large loop regime.

We start by confirming that, as shown in \cite{Sousa:2016ggw}, it is crucial to model the gravitational wave contributions of strings heavier than the fundamental string, as they give rise to additional structure in the spectrum. Since the cutoff frequency of GW emission of a given string species, $i$, is given by\footnote{Recall we have the loop size given in terms of the time of chopping by $l_i=\alpha_i t$, and the fundamental frequency of the GW it emits is given by $f_i=2/l_i$.} $f_{\rm min,i}=2/(\alpha_i t_0)$, and since $\alpha_i$ can vary between string species, the cut-off frequency becomes species-dependent. If $\alpha_i$ differs between species, this leads to regions in frequency space where the GW emission has stopped for fundamental strings, but continues for heavier strings, as discussed earlier (cf. low frequency regions of Fig. \ref{fig:smallLoops} corresponding to small loops with $\alpha_i=2\Gamma G\mu_i$) for $g_s=0.04$ and $g_s=0.3$. The result is a multi-peaked spectrum at low frequencies -- a clear signature of the GWs arising from a multi-tension network. 

However, this is not the whole story. In addition to multiple peaks at low frequencies, there can also be areas in frequency space where GW emission from heavier strings dominates over emission from fundamental strings, even though the latter is not negligible (see leftmost plot of Fig. \ref{fig:bigLoops} corresponding to large loops with $\alpha_{L,i}=0.37$ for $g_s=0.04$). As the birth times of loops relevant to a given GW frequency (Eq. \ref{eq:tb}) depend on the tension of the string species (and loop density is cut off when $t<t_b$), the integral in Eq. (\ref{eq:omegaj}) can be smaller for the fundamental strings than for heavier strings, despite the fact that the smaller correlation lengths result in many more loops of fundamental string overall. This effect is distinct from the appearance of multiple peaks (which is a direct consequence of having a hierarchy of loop sizes) and is present even when $\alpha_i$ is the same for all species.

In both cases, however, GW contributions from heavier string species are strongest for low values of the string coupling $g_s$ (for which the hierarchy in string tensions is higher) and show up at the low frequency part of the spectrum.

The changes in VOS parameters shown in Fig. \ref{fig:3vs7} also translate to noticeable differences in the GW background spectra between the 3-string and 7-string models. In all cases, except at very large string couplings, including heavier ($i>3$) strings in the VOS equations significantly boosts the GW contributions from the D and FD-strings. In the small-loop limit, this can be seen as mainly an enhancement in the secondary peak and in the large-loop limit as a general increase in magnitude. Since the modifications to the VOS parameters were largest at small couplings, the changes in the GW spectra are also most visible at low string coupling values. 

Our results, summarised in Figs.~\ref{fig:smallLoops} and \ref{fig:bigLoops} show that, in both the small loop and large loop limits, there are distinct signatures in the GW emission from heavier string species. These are more prominent for small $g_s$ as this gives rise to both a larger tension (leading to higher emission amplitude) and a larger loop size (leading to a lower frequency peak) for the heavy D and FD strings. These results are in agreement with reference \cite{Sousa:2016ggw}, up to changes in the locations of the peaks caused by different choices/interpretations of the parameter $\alpha_i$, which in the case of reference \cite{Sousa:2016ggw} corresponds to $\alpha_{L,i}$ in our notation. Thus, the physical loop sizes differ (between this work and reference \cite{Sousa:2016ggw}) by a factor of twice the time-normalised correlation length, $2\xi_i$, in the small loop limit, and by a factor of 3.7 in the large loop limit. Other, smaller, differences arise from our more accurate modelling of the cosmic superstring network achieved by (a) including decay processes of heavier strings down to the three lightest species (as discussed above), and (b) taking into account the dependence of the VOS parameters $\Tilde{c_i}$, $\Tilde{d}^k_{ij}$ on the scaling velocities of each string species.   

\subsection{Fixing model parameters}
\label{sec:parameters}

We have demonstrated that it is necessary to include several species of heavier strings when solving the VOS equations but not necessary to calculate the gravitational wave contributions of more than the three lightest string species. Before we proceed with fitting the model to NG15 data, we need to discuss the remaining model parameters $\{\Gamma,B,w,q,\alpha_{i}\}$. Since we will already be fitting at least a 2D space of fundamental string tension and string coupling, it is computationally unfeasible to keep all of the above as free parameters and most will need to be fixed at physically motivated values.

\subsubsection{GW emission power, $\Gamma$}
\label{sec:gamma}

$\Gamma$ gives the total GW power emitted by string loops over the course of their life (in units of $G\mu$) and has usually been considered as a constant of order $\mathcal{O}(50)$. Recent numerical simulations exploring the effect of gravitational back-reactions on Nambu-Goto cosmic string loops have found that $\Gamma$ may vary with time \cite{wachter_more_2024}. The potential effect of a time-evolving $\Gamma$ on the gravitational wave background is to suppress the signal with up to a $\sim 30\%$ decrease at high frequencies. This paper mainly focuses on the low-frequency end of the GWB spectrum, as that is the domain in which the potential signature of heavier string species is clearest and where NANOGrav has found evidence for the SGWB. As such, we will fix $\Gamma=50$ for all string species.

\subsubsection{Loop size, $\alpha_i$}\label{sec:loopsize}

One of the most important parameters in determining the shape and magnitude of the GW background spectrum is the size of string loops when they are chopped off from the string network. Motivated by the scaling behaviour of cosmic (super)string networks, it is commonly assumed that their size is a constant fraction of the string correlation length at the time of birth so that $l_i(t_b)=\alpha_i t_b$, where $\alpha_{i}$ is a constant. The exact value of $\alpha_i$ is unknown. Throughout the years, investigations into cosmic strings have shown loops produced at sizes comparable to the string thickness \cite{bevis_fitting_2008,vincent_correlations_1997}, sizes around the gravitational backreaction scale, $\alpha_i=2\Gamma G\mu_i$ \cite{bennett_high-resolution_1990,albrecht_evolution_1985,vincent_scaling_1997,allen_cosmic-string_1990}, and at scales much closer to the characteristic length of the string network \cite{olum_cosmic_2007,martins_fractal_2006,ringeval_cosmological_2007,vanchurin_scaling_2005,vanchurin_scaling_2006}. The most recent large-scale numerical simulation into the production of cosmic string loops finds a double-peaked spectrum of loop production, where around 90\% of network energy goes into loops at the gravitational backreaction scale and the remaining 10\% into large loops with $\alpha_i\sim 0.1$ \cite{Blanco-Pillado:2013qja}. Note in that paper the size of loops is expressed as a fraction of the horizon size during the radiation era, i.e $\alpha_h$, which they define as $\alpha$. This differs from the definition used in this work by a factor of 2.

Given the lack of a clear consensus, we will look at GWs emitted at the two scales found by \cite{Blanco-Pillado:2013qja} as these will allow us to look at both the large and small loop physics of GW emission. The results of \cite{Blanco-Pillado:2013qja} are obtained for radiation-era cosmic strings, which have a time-normalised correlation length of $\xi_r=0.27$, so that even in the large loop limit we have $\alpha_i<\xi_r$. As seen in Fig. \ref{fig:3vs7}, at small values of $g_s$, our correlation lengths can be as low as $\xi\sim0.03$. As it would be unphysical to assume that, on average, loops are born with a size greater than the correlation length, we will instead fix $\alpha_{L,i}=0.37$. This recreates $\alpha_{i}=\alpha_{L,i}\xi_i=0.1$ for ordinary cosmic strings and ensures that loop sizes stay below the correlation length throughout the parameter space for cosmic superstrings. In the small loop limit, we will still set $\alpha_i=2\Gamma G\mu_i$ in agreement with \citet{Blanco-Pillado:2013qja}. Note that it is not clear that results derived from cosmic string simulations can be directly applied to cosmic superstrings and $\alpha_i$ can, in principle, take any other value. Nevertheless, we would expect these scales to be representative of the range of possible values.

\subsubsection{Cusp-dominated vs kink-dominated emission}
\label{sec:cuspsVskinks}

Previous research into GW emission from cosmic strings and cosmic superstrings have mostly focused on emission from cuspy loops, especially bursts of GWs that are expected to be emitted in the vicinity of the cusps \cite{damour_gravitational_2001,damour_gravitational_2000,yonemaru_searching_2020,Damour:2004kw}. This may no longer be the optimal focus as recent simulations have shown that gravitational backreaction during loop evolution may smooth out the loop structure and prevent the formation of cusps \cite{wachter_numerical_2024}, which would then imply the leading source of GW emission are from kinks propagating around loops. For earlier work pointing to the importance of kinks in the GW signal from cosmic (super)strings see \cite{Binetruy:2010cc}.

The loop structure plays a role when summing over all the harmonic modes at which string loops emit GWs, as seen in Eq.~(\ref{eq:Omegajsum}). Including extra harmonics broadens the emission peaks when compared to the principal harmonic seen in Figs. \ref{fig:smallLoops} and \ref{fig:bigLoops} and leads to a near-flat high-frequency plateau. The transition between the peak of GW emission and the plateau is sharper for kink-dominated emission due to a larger damping of harmonic modes in Eq.~(\ref{eq:Omegajsum}) (recall for kinks $q=5/3$ whereas for cusps $q=4/3$). Given there is still uncertainty in the distribution of cusps and kinks on a typical loop, we will model GW emission from both cuspy and kinky loops when fitting to the data.

\subsubsection{Size of the extra dimensions and energy of the loops}
\label{sec:otherParams}

We have now seen how variations in loop size, number of string species and loop structure affect GW emission. These are the ``main" parameters, as varying them leads to the largest changes in the resulting GW spectrum. The final two parameters that need to be fixed before proceeding to fit the model to data are the fraction of junction energy transferred to loops during string intersection, $B$, and the parameter $w$, which describes the effective size of the extra dimensions.

As discussed in section \ref{sec:model}, the volume parameter $w$ is the ratio between the minimum volume and effective volume occupied by the fundamental string in the compact extra dimensions. A small value of $w$ corresponds to a larger volume of the compact dimensions being available to the strings, while $w=1$ corresponds to compactification at the string scale (or to highly warped extra dimensions) so that volume effects from the extra dimensions become negligible. Note that the effective volume is only a lower bound to the total volume of the extra compact dimensions. As more of the extra dimensions become traversable to the strings, the larger the probability of the strings ``missing" each other in the extra dimensions when intersecting. This leads to a decrease in intercommuting probability, corresponding to an increase in the number of string loops and a boost to the GW signal. Since the heavier strings oscillate less in the extra dimensions, the effective volume they explore is much smaller than for the fundamental strings and the boost in GW magnitude is much smaller. 

\begin{figure*}
    \centering  
    \includegraphics[width=0.95\textwidth]{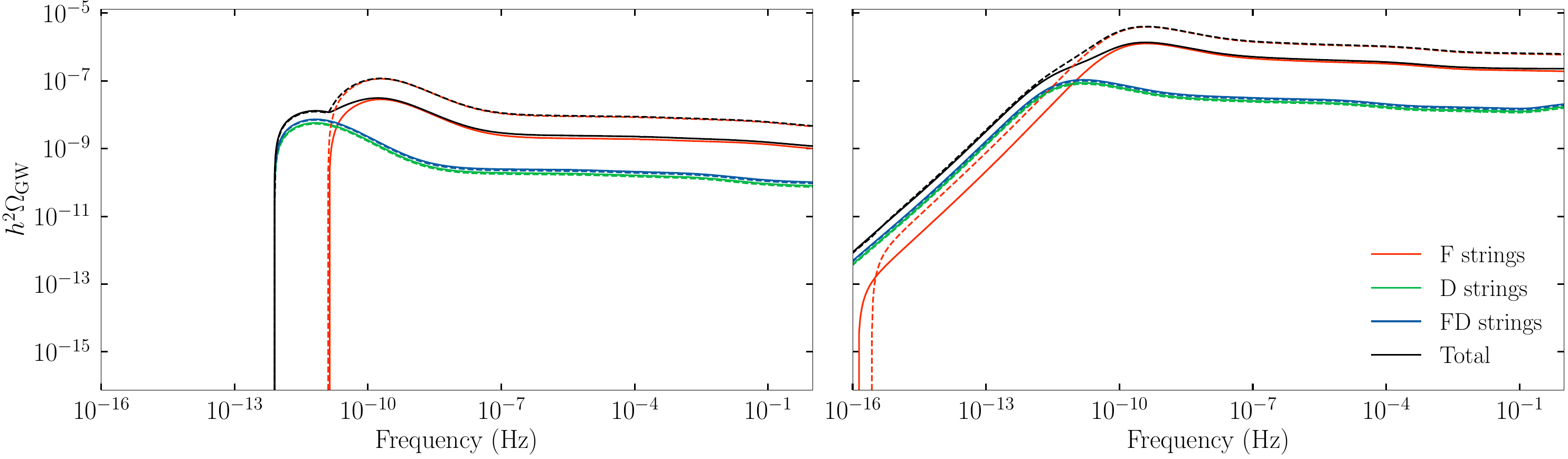}
    \caption{Plots of the gravitational wave background spectra generated at different sizes of extra dimensions -- w=0.1 (dashed lines) and w=1 (solid lines). The string tension is set at $G\mu_1=10^{-9}$ and loop sizes are $\alpha_i=2\Gamma G\mu_i$ on the left and $\alpha_i=0.1$ on the right. Only the fundamental mode is shown.}
    \label{fig:w} 
\end{figure*}

Comparison spectra for $w=0.1$ and $w=1$ can be seen in Fig. \ref{fig:w}. As varying $w$ has a strong impact on the resulting spectra and the value of $w$ is physically interesting, encoding information on the size/warping of the extra dimensions, we will treat this as a free parameter.

Increasing $B$ leads to additional kinetic energy in string loops and, hence, a higher velocity. This, in turn, corresponds to a larger correlation length and a smaller GW signal. The effect is strongest for heavier string species at large string coupling $g_s$, but in general, it is minor enough to be ignored. For ease of computation, we fix 
$B$ at a constant value, choosing $B=1$ as then there is explicit energy conservation within our system.

\section{Results and Discussion}
\label{sec:results}

Out of the set of parameters determining the cosmic superstring GW spectra, $\{\Gamma,B,\alpha_i,q,\mu_1,g_s,w\}$, we have fixed the values of $\Gamma$ and $B$, chosen binary scales for $\alpha_i$ and $q$ and will treat $w,\mu_1$ and $g_s$ as free parameters. We fit our model to the most recent PTA stochastic gravitational wave background dataset -- the NANOGrav 15-year dataset \cite{NANOGrav:2023gor}. To accurately model the high-frequency behaviour of the cosmic superstring SGWB spectrum, $j_{\rm max}=10^{12}$ harmonic modes are summed to produce each spectrum. To quantify the quality of our fit, we compute likelihoods following \citet{Ellis:2023tsl} as
\bea
    \mathcal{L}=\prod_{m=1}^{14} P_m(\Omega_{\rm GW}(f_m)),
\eea
where for each frequency bin $m$, $P_m(\Omega)$ are the posterior probability density functions of the NG15 Hellings-Downs-correlated spectrum analysis and $\Omega_{\rm GW}(f_m)$ is the cosmic superstring background spectrum energy density in that frequency bin. 

\begin{figure*}
    \centering  

    \includegraphics[width=0.95\textwidth]{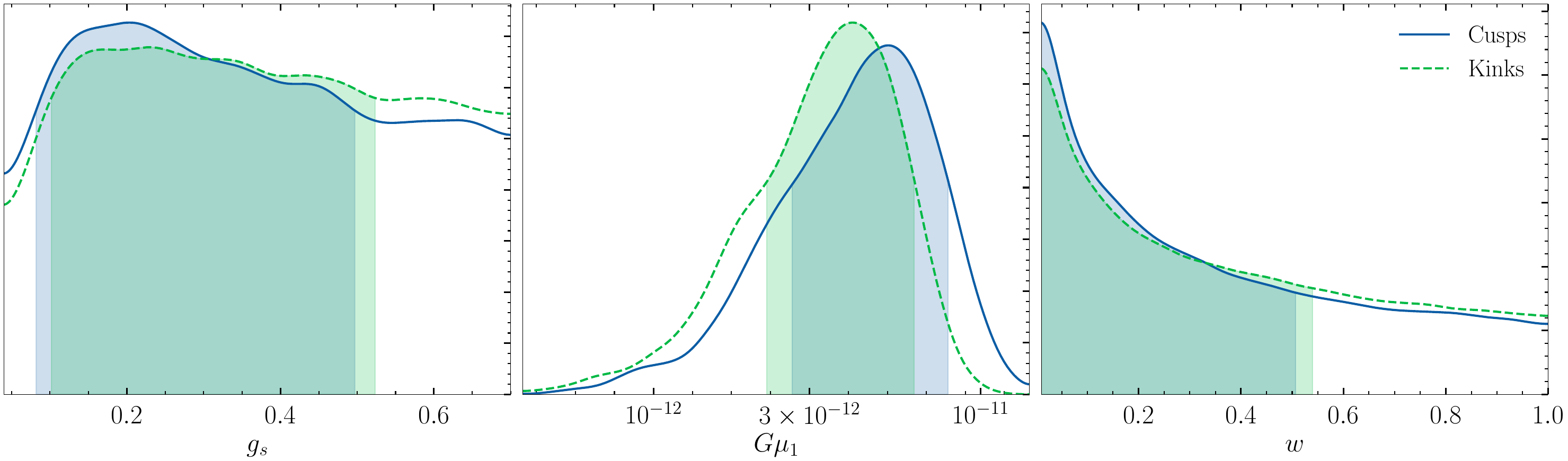}

    \caption{Likelihood posteriors for the large loop fit to the NG15 data. The solid (dashed) lines correspond to cuspy (kinky) loops.}
    \label{fig:largeLoopMarginals} 
\end{figure*}

\begin{figure*}
    \centering  
    
        \includegraphics[width=0.95\textwidth]{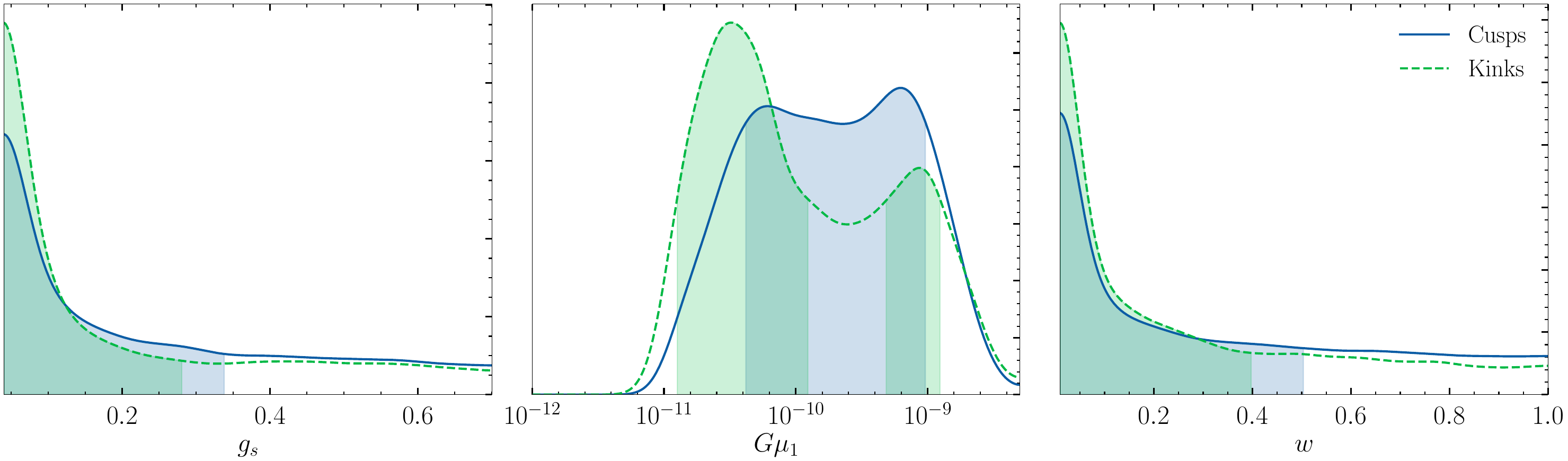}

    \caption{Likelihood posteriors for the small loop fit to the NG15 data. The solid (dashed) lines correspond to cuspy (kinky) loops.}
    \label{fig:smallLoopMarginals} 
\end{figure*}

In total, we produce four 3-dimensional likelihood contours -- large and small, cuspy and kinky loops. Note that here we assume all loops are formed at a given constant $\alpha_i$. This is different from the assumptions made by NANOGrav when producing their cosmic superstring fit \cite{NANOGrav:2023hvm}. These differences and their effects will be discussed in Sec. \ref{sec:NGmodel}.

\begin{table}
\centering
\resizebox{\columnwidth}{!}{%
\small  
\begin{tabular}{cccccc}
    \hline 
    \hline  
    Model & $-2\ln \mathcal{L}_{\rm max}$ & $\log_{10}(G\mu_1)$ & $g_s$ & $w$\\
    \hline
    \hline
    Cuspy big loops & $47.9$ & $-11.4^{+0.3}_{-0.2}$ & $<0.69$ & $<0.90$\\
    & {\scriptsize } & {\scriptsize -11.8} & {\scriptsize 0.22} & {\scriptsize 0.01} \\
    \hline
    Kinky big loops & $47.8$ & $-11.5^{+0.3}_{-0.2}$ & $<0.70$ & $<0.91$\\
    & {\scriptsize } & {\scriptsize -11.9} & {\scriptsize 0.26} & {\scriptsize 0.01} \\
    \hline
    Cuspy small loops & $53.6$ & $-9.7_{-0.7}^{+0.7}$ & $<0.63$ & $<0.88$\\[-1pt]
    & {\scriptsize } & {\scriptsize -10.8} & {\scriptsize 0.04} & {\scriptsize 0.01}\\
    \hline
    Kinky small loops & $53.5$ & $-9.9_{-0.5}^{+1.0}$ & $<0.61$ & $<0.83$\\[-1pt]
    & {\scriptsize } & {\scriptsize -10.9} & {\scriptsize 0.05} & {\scriptsize 0.01} \\
    \hline
    \vspace{1pt}
\end{tabular} }
\caption{The log-likelihoods and posterior mean values (with 68\% confidence intervals) for the models fitted to NG15. Upper $2\sigma$ limits are given for $w$ and $g_s$. The best-fit values for the multidimensional fits are given in small print below the posterior values.}
\label{tab:parameters}
\end{table}

\begin{figure*}
    \centering  
    \includegraphics[width=0.95\textwidth]{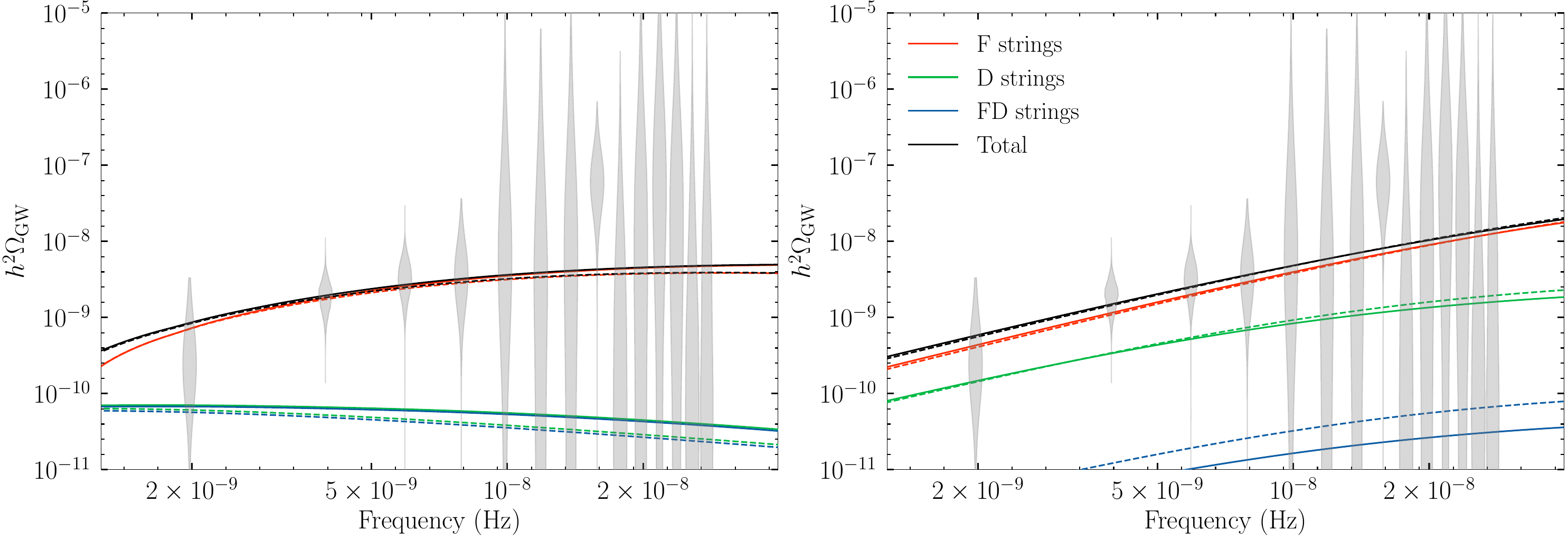}

    \caption{The best-fit spectra in the small (left) and large (right) loop cases. The solid (dashed) lines correspond to cuspy (kinky) loops. The NANOGrav 15-year data is shown as violins.}
    \label{fig:bestfits} 
\end{figure*}

The marginalized posterior fits in all three free parameters, along with their 68\% confidence regions, can be seen in Fig. \ref{fig:largeLoopMarginals} for large loops and in Fig. \ref{fig:smallLoopMarginals} for small loops. The parameter ranges, along with the parameters corresponding to the best-fit points, can be seen in Table \ref{tab:parameters}. The spectra corresponding to the best-fit points (along with the NG15 PDFs) can be seen in Fig. \ref{fig:bestfits}. 

\subsection{Likelihood Posteriors}

As can be seen in Fig.~\ref{fig:bestfits} and Table~\ref{tab:parameters}, both large and small loops fit well with the NG15 data with large loops being favoured by the data. There is no major difference between the results obtained from cusp and kink-dominated emission. This is not unexpected, as the data is in the nanohertz range, whilst the change in loop structure mainly impacts higher frequencies (higher harmonics).

The most well-constrained parameter in all cases is $\log(G\mu_1)$. The posterior is sharply peaked as the peak frequency and amplitude of GWs depend strongly on the tension. Fitting both the low-frequency downturn and amplitude of the NG15 data requires a precise value for the tension.

Both the marginals for $g_s$ and $w$ peak at lower values. For large loops, these peaks are wide and the marginals are overall relatively flat. For small loops, the peaks are sharp but also narrow, resulting in a wide range of allowed values. As can be seen from Table. \ref{tab:parameters}, all models have nearly the same $2\sigma$ upper limits, with approximately $g_s<0.65$ and $w<0.9$.

\subsection{Likelihood comparisons}

The approximate $\chi^2$($\approx -2\ln \mathcal{L}_{\rm max}$) values for the best-fit spectra can be seen in Table \ref{tab:parameters}. We see that the best fit is given by kinky big loops, with approximately the same likelihood as cuspy big loops. Small loops give worse fits to the data with $\Delta \chi^2\approx7$ when compared to the large loop fits. To compare the quality of fits with other models we use the Bayesian information criterion, which in our case is given by
\bea
    \text{\rm BIC}=k\ln 14-2\ln \mathcal{L}_{\rm max},
\eea
where $\mathcal{L}_{\rm max}$ is the maximum (best-fit) likelihood and $k$ is the number of parameters estimated by the model.

We start by comparing our results with the (cuspy) cosmic string best-fit obtained by the NANOGrav collaboration ($\text{BIC}_{\rm NG}^{\rm cusp}$) \cite{NANOGrav:2023hvm}. We obtain $\text{BIC}_{\rm NG}^{\rm cusp}=76$ and $\text{BIC}^{\rm cusp}=56$ giving $\Delta\text{BIC}^{\rm cusp}=-20$, which signals a much better fit. This is to be expected, as ordinary cosmic strings (as opposed to cosmic superstrings) are known to have a spectrum that is too flat to explain the NANOGrav SGWB measurements.

It is much more informative to compare the cosmic superstring results with current up-to-date models of supermassive black hole binary mergers, as they are the leading astrophysical candidate for the source of the SGWB. The three black hole binary models investigated in \citet{Raidal:2024odr} -- circular binaries evolving purely gravitationally, eccentric binaries evolving gravitationally and circular environmentally driven binaries -- have $\text{BIC}_{\rm SMBH}^{\rm circ}=60$, $\text{BIC}_{\rm SMBH}^{\rm ecc}=50$, $\text{BIC}_{\rm SMBH}^{\rm env}=57$. We see that the large loop fits are comparable in likelihood to the environmental model and slightly worse than the eccentric model, while the small loop fits are comparable to the circular binary model. This shows that cosmic superstring models are competitive with the best current astrophysical models.

\subsection{Comparison to NANOGrav fit}
\label{sec:NGmodel}

\begin{figure*}
    \centering  

    \includegraphics[width=0.95\textwidth]{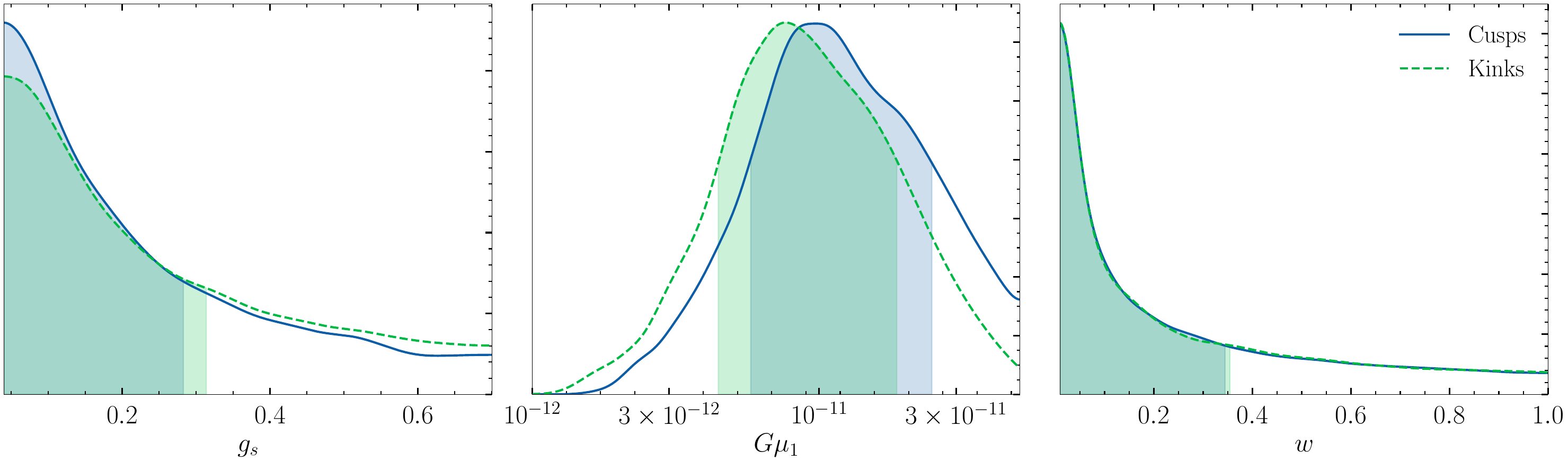}

    \caption{Likelihood posteriors obtained from fitting the model considering $10\%$ large loops and $90\%$ non-contributing small loops to the NG15 data. The solid (dashed) lines correspond to cuspy (kinky) loops.}
    \label{fig:NGmarginals} 
\end{figure*}

When analysing their data, the NANOGrav collaboration used a simplified cosmic superstring model, where only one type of string dominates \cite{NANOGrav:2023hvm}. In this case, the treatment amounts to a model of cosmic strings with a reduced intercommuting probability $P$, assumed to rescale the spectrum as follows:
\bea
    \Omega_{\rm SUPER}(f)=\frac{1}{P}\Omega_{\rm CS}(f),
\eea
where $\Omega_{\rm CS}$ is the spectrum obtained from cosmic strings at a given tension and $P$ is treated as a constant free parameter\footnote{The scaling $\Omega_{\rm SUPER}(f)\propto P^{-1}$ can be understood in terms of the VOS model as taking the loop chopping efficiency $\tilde c\propto P^{1/2}$ leading to a scaling $L\propto P^{1/2}$ in the correlation length \cite{Sakellariadou:1990nd,Dvali:2003zj} and a corresponding boost in the string density $\rho=\mu/L^2\propto P^{-1}$.}.
There are two main downsides to this approach. First, as shown in this paper, considering heavier string species in the cosmic superstring network is crucial to accurately model the resulting GW spectrum across its full frequency range. In particular, cosmic superstring networks have characteristic features (the presence of multiple peaks and/or domination of heavy string species at low frequencies) and so modelling the GW signal through a single string network could lead to missed opportunities in potentially discovering these exciting features. However, it turns out that, 
within the NG15 frequency range, the fundamental string does in fact dominate and the one-string approximation is consistent with the data, as can be seen in Fig. \ref{fig:bestfits} from the negligible contributions of the D and FD-strings. The second and main downside of the single-string approximation comes from the unclear physical interpretation of the parameter $P$. Although the one-string approximation can provide a best-fit likelihood and tension, the fundamental string parameters $g_s$ and $w$ are in fact encoded in $P$ and information about them can only be obtained through the full treatment outlined in this paper.

In comparing our results to the one-string approximation within the NG15 frequency range, as well as finding the parameter values for $g_s$ and $w$ for the NANOGrav fit, we have to make a slight modification to the model that has been used so far. NANOGrav assumes that $90\%$ of loops are formed at a small size and are redshifted away without contributing to the GW spectrum. The remaining $10\%$ of loops are large and become the dominant contributors to the SGWB. This corresponds to inserting a factor of $0.1$ to the loop number density of our large loop model. 

Performing this analysis gives cuspy (kinky) likelihood posteriors with 68\% confidence intervals of $\log_{10}(G\mu_1)^{NG}=-11.0^{+0.4}_{-0.2}$($-11.1^{+0.4}_{-0.3}$), and $2\sigma$ upper limits of $g_s^{NG} < 0.59$($<0.61$) and $w^{NG} < 0.84$($<0.84$). The posterior likelihood marginals obtained can be seen if Fig.~\ref{fig:NGmarginals}. Note that the 68\% interval on the string tension is significantly different to that found by \citet{NANOGrav:2023hvm} (confidence interval $[-12.08,-11.50]$) and \citet{Figueroa:2023zhu} ($\log_{10}(G\mu_1)^{NG}=-11.79^{+0.31}_{-0.19}$), as we are using smaller loop sizes in accordance with our more accurate modelling of the string correlation lengths.

This discrepancy highlights another shortcoming of the $1/P$ modelling approach: its failure to accurately account for variations in correlation length. By calculating the ordinary cosmic string GW signal and then rescaling it by a factor of $1/P$, this approach, in effect, takes the correlation length to scale as\footnote{Also note that, while the scaling $L\propto \sqrt{P}$ is consistent with flat space simulations \cite{Sakellariadou:1990nd} and early theoretical expectations \cite{Dvali:2003zj}, string evolution simulations in an expanding universe find a weaker dependence, $L\propto P^{1/3}$, in both the matter and radiation eras \cite{Avgoustidis:2005nv}.} $L\propto \sqrt{P}$ without a corresponding rescaling of the loop size, which is fixed to $\alpha=0.1$.
This is relevant in the cosmic superstring case, as discussed in section~\ref{sec:loopsize}, where fixing $\alpha_i=0.1$ can lead to loops with lengths multiple times longer than the correlation length of each string species.
In fact, when we impose $\alpha_i = 0.1$ in our analysis (along with the $0.1$ factor in the loop number density) we recover tension bounds close to those in \cite{NANOGrav:2023hvm,Figueroa:2023zhu}, and reproduce NANOGrav’s best-fit spectrum exactly. The best-fit tension of $log_{10}(G\mu_1)=-11.94$, however, corresponds to $g_s=0.06$ and $w=0.01$, implying a fundamental string correlation length of $\mathcal{O}(0.01)$, as shown in Fig.~\ref{fig:3vs7}. At such small correlation lengths, the loop sizes used to obtain this spectrum become unphysically large.

\subsection{Forecast for future experiments}

\begin{figure*}
    \centering  
    \includegraphics[width=0.95\textwidth]{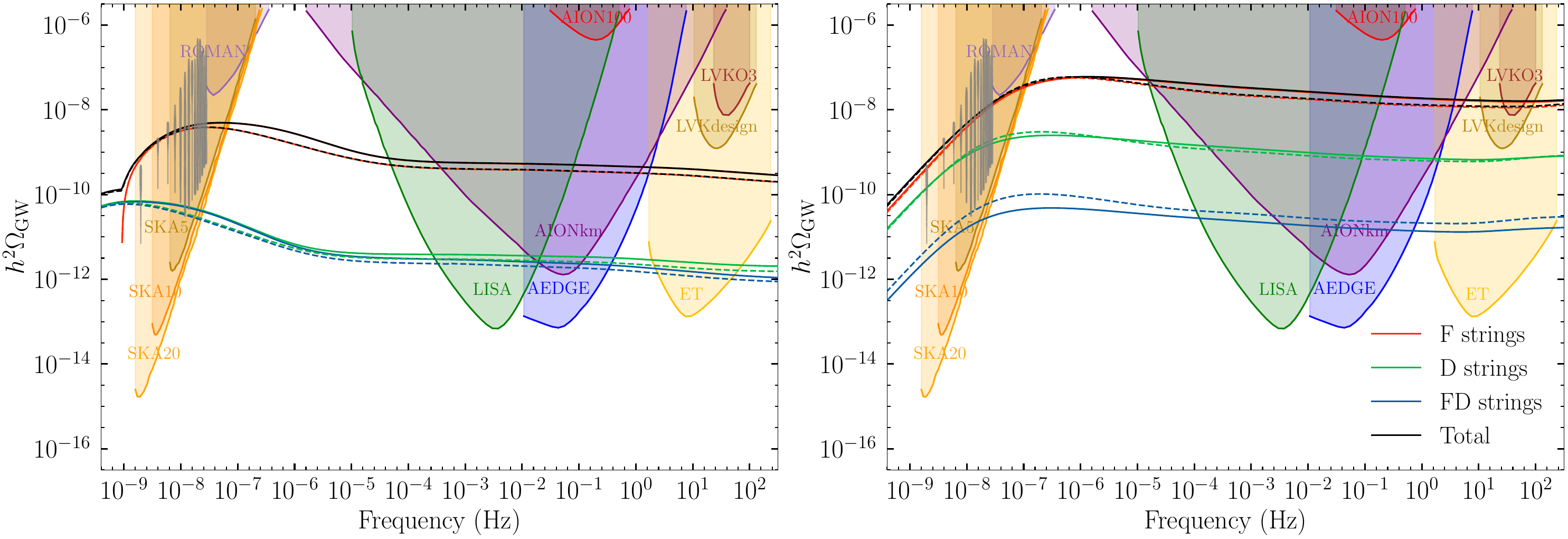}

    \caption{The best-fit spectra in the small (left) and large (right) loop cases. The solid (dashed) lines correspond to cuspy (kinky) loops. The NANOGrav 15-year data is shown as violins, and the sensitivity curves of current and (projected) future experiments, taken from \cite{Ellis:2023tsl}, are shown by shaded regions.}
    \label{fig:forecast} 
\end{figure*}

As shown by the best-fit spectra in Fig. \ref{fig:bestfits}, within the frequency range probed by NANOGrav, there are no signatures unique to GWs emitted by cosmic superstrings.  It is also difficult to separate between cuspy and kinky spectra, as their differences only become apparent at high frequencies. To see how the full spectrum of cosmic superstring gravitational waves can be probed by current and future experiments, we plot the small and large loop best-fit spectra along with the sensitivity curves of current and future experiments, taken from \cite{Ellis:2023tsl}, in Fig. \ref{fig:forecast}.

We start by noting that cosmic strings and cosmic superstrings, unlike astrophysical sources, emit gravitational waves over a very wide frequency range. This, more than any specific features of the spectrum, can be used to determine whether the SGWB measured by current and future GW experiments originates from astrophysical or cosmological sources.

Gravitational waves in the nanohertz range are currently being probed by NANOGrav \cite{NANOGrav:2023gor}, with low-frequency ranges beyond current PTAs being within the sensitivity ranges of SKA \cite{Janssen:2014dka} and ROMAN \cite{Wang:2022sxn}. The microhertz to hertz range will be covered by LISA \cite{Bartolo:2016ami,Caprini_2019,LISACosmologyWorkingGroup:2022jok} and cold-atom-based experiments AEDGE \cite{AEDGE:2019nxb} and AION \cite{Badurina:2019hst}. This range could be useful in differentiating between kink and cusp sourced GW emission, as the fits to NG15 data differ significantly in magnitude within this range. Finally, the highest frequency range ($\sim 10-100$ Hz) of the spectrum could already be within the sensitivity range of LIGO/Virgo/KAGRA \cite{LIGOScientific:2014pky,LIGOScientific:2016fpe,LIGOScientific:2019vic}(especially when they reach their maximal sensitivities), although drop-offs in magnitude could be caused by modifications to standard cosmic evolution \cite{Ellis:2023tsl, Datta:2024bqp}. Similar reductions in high-frequency magnitude could also be caused by explicit modelling of gravitational back-reaction on cosmic superstring loops \cite{wachter_more_2024}. Due to its larger magnitude at high frequencies, the large-loop spectrum will be easiest to probe by current and proposed future experiments. As the small-loop spectrum has a sharper drop from peak emission intensity, its high-frequency tail is beyond the reach of LIGO/Virgo/KAGRA (but within the projected sensitivity of the Einstein Telescope \cite{Hild:2010id,Punturo:2010zz}).

The clearest signatures of cosmic superstrings in the SGWB spectrum appear to be from the heavy string contributions seen at low frequencies. The lowest frequencies in gravitational waves are probed by PTA experiments (such as SKA and NANOGrav), with the lowest frequency being limited by observational time. 
Probing the earliest ``smoking gun" signature, the sharp transition between emission peaks seen at $\sim 1 \rm nHz$ in the small-loop best-fit, requires around $30$ years of observational time. Measuring any peaks and structures at even lower frequencies are unfortunately outside the feasibility range of current experimental methods. 

\section{Conclusions}
\label{sec:conclusions}

We have presented the most up-to-date model of cosmic superstrings, their evolution and the stochastic gravitational wave background they source. We have shown that the accurate modelling of their evolution  requires the inclusion of both the lightest strings (F and D-strings) and several of their heavier bound states. These heavier states interact with the lighter strings (and between themselves) and decay into the lighter states. This adds additional energy to the networks of lighter strings, boosting their root-mean-square velocity and reducing their average correlation lengths. The important role played by heavier strings has also been highlighted in the context of determining CMB anisotropies in \cite{Pourtsidou:2010gu,Charnock:2016nzm}. 

Full modelling of the stochastic gravitational wave background (SGWB) generated by cosmic superstring networks also requires the inclusion of heavier strings. Although at higher frequencies the fundamental string dominates the GW energy density, at lower frequencies 
there are regions in parameter space where heavier strings contribute significantly to the signal and eventually dominate over the lightest string GW contribution. This can be understood in terms of the different correlation lengths, loop sizes, and tensions of each string species. The signatures of heavier strings seen at low frequencies depend on the strength of the string coupling $g_s$, and the size of string loops $\alpha_i$. If cosmic superstring loops are large when chopped from the string network, the low-frequency tail of the SGWB is smooth, with the heavier strings eventually overtaking the fundamental string in spectrum magnitude. When the loops are born at a small size, the SGWB drops off sharply at low frequencies, forming a distinct peak for the emission from heavier species. This peak is shifted towards lower frequencies for heavier strings, leading to a distinct multi-peaked spectrum, as clearly seen in the leftmost plot of Fig.~\ref{fig:smallLoops}.

Fitting the cosmic superstring model to the NANOGrav 15-year gravitational wave background data gives well-constrained values for the fundamental string tension at both small and large loop sizes. As string loops oscillate, the dominant fraction of gravitational waves can come from either cusps (single points moving instantaneously at highly relativistic speeds) or kinks (discontinuities in the loop tangent vector). Within the NANOGrav frequency range, this origin does not have a significant effect on the best-fit spectra or parameters, as we obtain expectation values for the fundamental string tension of $\log_{10}(G\mu_1)=-11.4^{+0.3}_{-0.2}$($-11.5^{+0.3}_{-0.2}$) for gravitational waves originating from large cuspy (kinky) cosmic superstring loops and $\log_{10}(G\mu_1)=-9.7^{+0.7}_{-0.7}$($-9.9^{+1.0}_{-0.5}$) for small cuspy (kinky) loops. The posteriors for string coupling, $g_s$, and $w$, which describes the size of extra compact dimensions, are much flatter and we set $2\sigma$ confidence bounds as $g_s<0.69$($<0.70$), $w<0.90$($<0.91$) for the large cuspy (kinky) loops and $g_s<0.63$($<0.61$), $w<0.88$($<0.83$) for small cuspy (kinky) loops.

Finally, we repeat the analysis by assuming the string loops consist of 10\% large loops and 90\% small loops, where only large loops contribute to the GW signal. In this case, we obtain bounds for the string tension of $log_{10}(G\mu_1)^{NG}=-11.0^{+0.4}_{-0.2}$($-11.1^{+0.4}_{-0.3}$) which are different from the bounds obtained by \citet{NANOGrav:2023hvm} and \citet{Figueroa:2023zhu}, who make the same assumptions on the string loop distribution.
These differences can be attributed to our modelling of the cosmic superstring network in terms of multiple correlation lengths, as opposed to the single-string approximation used in \cite{NANOGrav:2023hvm,Figueroa:2023zhu}. As discussed in section~\ref{sec:NGmodel}, calculating the superstring GW spectrum by rescaling an ordinary cosmic string spectrum can lead to loop lengths larger than the string correlation length.
We also place $2\sigma$ bounds on $g_s<0.59$$(<0.61)$ and $w<0.84$($<0.84$), which is not possible with the single-string approximation used in \cite{NANOGrav:2023hvm,Figueroa:2023zhu}. In all cases, the quality of fits is comparable to the current best astrophysical models involving supermassive black hole binaries, although large loops are favoured over small loops.

Extending the SGWB fitted to the NANOGrav data over the full range of GW emission frequencies shows that, at higher frequencies, both the large and small-loop fits can be easily probed by projected future gravitational wave experiments, such as LISA and the Einstein Telescope. These measurements can also be used to differentiate between cusp and kink-dominated emission, as the small-scale structure of loops affects the damping of higher harmonics (higher frequencies). Unfortunately, the low-frequency tail of the SGWB, which contains the clearest signatures of heavy cosmic superstring species, is out of reach of current experimental methods.

There remain significant uncertainties in modelling the GW signal from cosmic (super)string networks. The most important unknown is the size of string loops, which determines both the magnitude and shape of the spectrum as seen, for example, in Figs. \ref{fig:smallLoops} and \ref{fig:bigLoops}. This is a challenging and important open problem in the field. The best we can do at present is model various options, through different choices of loop distributions, or the parameter $\alpha$ as we have done in this paper. For cosmic superstrings, in particular, there are additional uncertainties arising from approximations/assumptions in the computation of the self- and cross-string interaction parameters $\Tilde{c}_i$ and $\Tilde{d}^k_{ij}$ in the multi-tension VOS model, the impact of compactification choices for the extra dimensions, and other assumptions relating to the formation and evolution of string junctions (see for example \cite{Rybak:2018mnl}). Nevertheless, our current modelling techniques provide a systematic way of quantifying these effects as we have exemplified in this paper. Indeed, given these modelling uncertainties, the relative robustness of the predicted GW signal is somewhat surprising. In particular, a single species toy-model can provide a reasonable approximation to the GW spectrum over most of the relevant frequency range, as heavier strings only contribute significantly at low frequencies. If the single-string approximation is made, however, it is important to directly model the dependence of correlation length on intercommuting probability in order to avoid unphysical loop sizes. 

The detailed modelling we have discussed here, crucially, links the GW spectrum with the underlying fundamental parameters of the model -- the tension of the fundamental string $\mu_F$, the string coupling constant $g_s$ and the size of the extra dimensions through $w$ -- and makes the characteristic features of cosmic superstring networks apparent at low frequencies. This will be important for constraining or detecting these effects with future GW data.

\begin{acknowledgments}

The work of AA, EJC and AM was supported by an STFC Consolidated Grant [Grant No. ST/X000672/1]. JR was supported by an STFC studentship [Grant No.\ ST/Y509437/1]. We are grateful to Lara Sousa and Ivan Rybak for helpful discussions. For the purpose of open access, the authors have applied a CC BY public copyright license to any Author Accepted Manuscript version arising.

\end{acknowledgments}

\bibliography{refs}

\end{document}